# Grain size modification in the magnetocaloric and non-magnetocaloric transitions in $La_{0.5}Ca_{0.5}MnO_3$ probed by direct and indirect methods.

M. Quintero[1,2], S. Passanante[1,3], I. Irurzun[1,3], D. Goijman[1], G. Polla[1].

1 Departamento de física de la materia condensada, GIyA, GAIANN, Comisión Nacional de Energía Atómica, Buenos Aires, Argentina.

2 Escuela de ciencia y tecnología, Universidad Nacional de General San Martin, Buenos Aires, Argentina.

3 Departamento de física, Facultad de ciencias exactas y naturales, Universidad de Buenos Aires, Buenos Aires, Argentina.

**Abstract.**

The influence of the grain size in the magnetic properties of phase separated manganites is an important issue evidenced more than a decade ago. The formation of long range ordered phases is suppressed as the grain size decreases giving place to a metastable state instead of the ground state. In this work we present a study of the magnetocaloric effect in the prototypical manganite $La_{0.5}Ca_{0.5}MnO_3$ as function of the grain size. The differences obtained using direct and indirect methods are discussed in the framework of domain walls in the ferromagnetic phase of the system.

## Introduction

The discovery in 1997 of giant magnetocaloric effect (MCE) near room temperature in Gd based compounds[1] triggered a constant growth in the number of scientifc publications dedicated to the study of the mentioned effect. The main motivation is the high cost of production of Gd that difficult the production of magnetic refrigeration systems in commercial scale. A large number of compounds has been proposed to replace Gd, such as As based compounds[2], heusler alloys [3] and manganites [4].

The MCE in solid materials is produced by the magnetic entropy change induced when an external magnetic field is applied. In standard ferromagnetic systems an increase in the magnetic field reduces the magnetic entropy and, if the field is applied adiabatically, the lattice thermal entropy increases, giving rise in the sample temperature change.

In more complex systems the above simplified scenario may not be enough to describe the behavior of the entropy change. A strong coupling between different degrees of freedom (magnetic, electronic, etc.) is usually responsible for such a mixed change of the state of the system by the application of a magnetic field. Depending on the characteristics of the different degrees of freedom the corresponding terms in the first law of thermodynamic may increase the heat change. But it can also be compensated, leading in a reduction, the suppression or even the inversion of the temperature change (the so called inverse magnetocaloric effect IMCE[5]). Because of this reason, a large number of scientific works has been devoted to the understanding of the MCE in cases beyond the standard ferromagnetic systems[6].

The most commonly used methods to study MCE can be divided in two well distinguished groups, according to the physical quantity that is measured to take account the effect.

The direct methods are those where the temperature change or the heat exchanged with the environment is directly measured. Once determined any of these magnitudes, the total entropy change can be estimated in non adiabatic conditions.

In the indirect methods MCE is obtained through thermodynamic relations between the entropy and other measured magnitud, such as magnetization or resistivity[7]. The most accepted way to obtain the entropy change is using a Maxwell´s relation (MR)

$$\frac{\partial S}{\partial H} = \frac{\partial M}{\partial T}$$

Then, the entropy change can be estimated permorming a numerical integration of a set of magnetization loops at different temperatures as

$$\Delta S(T,H) = \frac{1}{\Delta T}\int_0^H [M(T+\Delta T, H') - M(T, H')]dH'$$

The main advantage of this approach is the use of a standard experimental technique to reach the entropy values, instead of a specific setup designed to measure the sample temperature change[8].

In early works has been demonstrated that, under certain circunstances, the results were consistent with those obtained by direct methods[8]. But the use of the MR in cases where the system is out of equilibrium can lead to a overestimation of MCE [9]. In the last few years, due to the increase of the complexity of the studied compounds, the validity of the MR approach has been revised by a growing part of the scientific community[6][10][11][12].

The continuous search for materials with large MCE stimulated further research in complex magnetic oxides[13], including mixed valence manganese based compounds, commonly named *manganites*. One of the most interesting properties of manganites is the spatial coexistence of regions with different magnetic ordering, the so called phase separation phenomena [14]. In systems with phase separation it is posible to tune the magnetic and structural properties by a variety of parameters such as electric and magnetic field, strain, doping, confinement and grain size [15][16][17].

In most of the cases of phase separation coexists an insulating antiferromagnetic charge ordered phase (CO) and a metallic ferromagnetic one (FM) [18]. One of the most studied systems with phase separation is $La_{0.5}Ca_{0.5}MnO_3$[19,20]. In this system the coexistence between the different magnetic phases can be controlled by external stimuli (radiation, electric field) or by the modification of synthesis parameters, particularly modifying the grains size (GS) in ceramic samples [16]. The increase in the GS favor the long range ordering of the charge ordered (CO) state over the ferromagnetic metallic (FM). The low temperature CO ground state of the system is strongly suppressed for small GS, and when it is increased the system can reach the CO state. The influence of GS in the MCE was recently studied in phase separated systems, revealing a complex scenario where the validity of the methods used to estimate the magnitude of the effect must be carefully revised [21,22].

In this work we present a study of MCE in the manganite $La_{0.5}Ca_{0.5}MnO_3$ which presents phase separation. The study will be performed as function of GS. We will compare results obtained from differential thermal analysis (DTA) and from indirect measurements with particular focus on the use of the MR relation. The hysteresis of the magnetization loops will be also analyzed and described in the framework of domain walls displacement and related with the differences observed between direct and indirect methods.

It has to be noted that the understanding of the phase separation in the $La_{0.5}Ca_{0.5}MnO_3$ system escapes to the aim of this work. We will assume the phase separated scenario acepted and widely discussed in previous works [16,23] and we will not deal neither with the origin of the phase separation nor with the possibilities of any alternative description.

**Experimental.**

Polycrystalline samples of $La_{0.5}Ca_{0.5}MnO_3$ were synthesized following a citrate/nitrate decomposition method using 99.9% purity reactants. To increase de grain size sub sequential thermal treatments have been performed to the samples as is descrypted in Levy et. al. [16]. The grain size of the samples was estimated from SEM microphotographs.

Magnetization measurements were made in a Quantum Design Versalab with the VSM and the heat capacity accessories. For the DTA measurements we used a home made system formed by two Cernox CX-1080-SD thermometers (manufactured by Lake Shore Cryotronics) on a Teflon piece to ensure thermal insulation between the sample and the reference thermometers. The reference used was a piece of alumina.

The whole system was mounted in a Versalab´s transport puck, allowing us to perform magnetization, Cp and DTA measurements in the same range of magnetic field and temperature.

**Results**

As it was previously reported, the change of the GS of the samples induce important changes in the magnetic behavior. We can see those changes in figure 1, where we show magnetization measurements of the entire set of samples with an applied field of 1 T on cooling. The grain size of the samples goes from 180 nm in sample A to 1300 nm in sample E (see table in figure 1 for details).

All the samples present an FM ordering at around the same temperature $T_c$ = 250 K but, while the sample with smallest GS (A) remains FM in all the temperature range below $T_c$, a clear FM to anti ferromagnetic transition is observed in the rest of the samples at T = 150K.

Measurements are performed with H = 1, which is enough to saturate the FM phase but not strong enough to induce a ferromagnetic fraction enlargement[24].

Because of that, the FM fraction at low temperature can be estimated as the ratio between the magnetization at 50K of the sample and the same value on the sample A (fully FM).

In the inset of figure 1 we show the FM fraction at low temperature as function of the GS, being close to 20 % in the sample with the largest GS. This change in the magnetic behavior can be interpreted as an evidence of the frustration of the CO state (associated with the AFM ordering) due to small GS. The localization of the charges implies the presence of a long range Jahn-Teller distortion that is suppressed by the disruptive change in the lattice due to the grain boundary [16]. Similar behavior has been reported in other compounds [22,25,26], indicating that the GS is an extra ingredient to take into account when the magnetic properties are studied.

To analyze how GS affects the MCE we used two independent methods to estimate the magnitude of $\Delta S$ and $\Delta T$.

In the first method we used isothermal magnetization curves and the above mentioned Maxwell´s relation to obtain the adiabatic entropy change due to the application of the magnetic field. In figure 2 we present the temperature dependence of the entropy change for the different samples with an applied magnetic field of 3T.

In all the samples we observe a negative peak close to Tc that can be associated to the PM to FM transition. The maximum entropy change remains almost constant at 2-3 J/kg-K for the entire series of samples.

An additional (positive) peak is observed at a lower temperature, around 150 K. The maximum entropy change in this peak increases as the grain size became larger. According with the magnetization data, this peak can be associated with the FM to CO transition.

The obtained value of $\Delta S$ for the sample E (largest GS) for H = 3 T is 10 J/kg-K, similar to the obtained for pure Gd around room temperature[27] and in other half doped manganites, such as $Pr_{0.5}Sr_{0.5}MnO_3$ [28] and $Nd_{0.5}Sr_{0.5}MnO_3$ [29] measured using the same method.

According with the presented data we can conclude that the MCE has been enhanced increasing the grain size, since an additional peak in the entropy change is observed and its magnitude is controlled by the GS.

To complete the picture we performed differential thermal analysis measurements, allowing us to determine the sample temperature change during the application of the magnetic field. In all the cases the sample was zero field cooled to the target temperature and then the field was applied with a constant rate of 200 Oe/sec. and the heat exchanged with the environment has been taken into account[30].

In figure 3 we present the adiabatic temperature change ($\Delta T_{AD}$) extracted from DTA measurements for samples A, C and E. A positive peak can be observed around 225 K. This is consistent with the expected behavior from the entropy change associated with the PM to FM transition. Surprisingly, we do not observe any peak related with the FM to CO transition. Is has to be noted that according with the entropy change values obtained from magnetization, the expected temperature change should be three times larger than the observed from the PM/FM transition.

Another important aspect to consider is the presence of hysteresis in the magnetization as function of magnetic field curves. To examine these feature in depth, we calculated the magnetic work (W) defined as the area enclosed between the curves obtained increasing and decreasing the magnetic field (between zero and 3 Tesla). In figure 4 we present W as function of temperature for all the measured samples.

In all the samples W is almost zero above the Curie temperature, indicating the absence of hysteresis in the paramagnetic phase. But when the FM phase is present we observe a strong relation between the temperature dependence of W and the grain size.

For smaller grain size (samples A, B and C) the magnetic work presents an increase on cooling remaining a constant value below 175 K. Samples D and E present a maximum at 175 K, decreasing its value and keeping constant below 75 K. It is interesting to note that in the temperature range in which W peak occurs coincides with the range where the CO phase appears.

The temperature behavior of W can be explained considering the Jiles-Atherton model [31] to describe the magnetization curves. In this model the hysteresis is produced by impedances to domain wall motion caused by pinning sites encountered by the domain walls as the move. Because of that the system at a given field H can not reach the global minimum energy state, giving place to a hysteretic magnetization loop.

The pinning sites could be grain boundary or any kind of inhomogeneities within a grain, for example tangles of dislocation and precipitates or nonmagnetic inclusions. The model consider that the domain walls are flexible so they do not only can move but the can also bend. When the domain walls bend while being held by a pinning site this results initially in a *reversible* change in the magnetization.

In our case the formation of the CO phase increase the amount of pinning centers in the sample, enhancing the reversible change in the magnetization.

On cooling, at 200 K the CO phase start a nucleation process, increasing the density of pinning sites in the material which is reflected in the increase of W. Once the CO phase is nucleated, the nuclei starts to growths in size, decreasing the amount of FM phase present. As consequence, W is reduced as the magnetic signal decreases.

The origin of the peak in the $\Delta S$ is the presence of a reversible component to the magnetization in this temperature region.

The energy associated with the magnetization difference is not exchanged with the environment because is used internally to bend the wall domain and recovered when the magnetic field is turned off. Because of that we did not observe a temperature change in the sample in this temperature region. The reversible nature of the bending of domain walls makes the entropy change calculated by Maxwell´s relation convertible in magnetic work and not in heat exchanged with the environment.

Conclusions:

In summary, we presented a study about the influence of the grain size in the magnetic and magnetocaloric properties of $La_{0.5}Ca_{0.5}MnO_3$. The system is characterized by two well distinguished magnetic transitions, a PI to FM one at 225 K and a FM to a phase separated CO + FM at 150 K. The MCE associated with the first transition do not present a significant dependence with the grain size and results extracted from magnetization measurements are in good agreement with those obtained from DTA measurements.

The second transition, related with the formation of the CO phase, presents a strong dependence with the grain size. The entropy change obtained from magnetization measurements is not consistent with the temperature change extracted from DTA measurements.

The hysteretic behavior in the magnetization loops that was explained using a Jiles and Atherton model of domain walls in the FM phase. In this framework the formation of the CO phase modifies the density of pinning sites increasing the hysteresis in the magnetization loops. This additional pinning sites increases the magnetic entropy calculated by Maxwell relation, but this entropy can not be used for applications since it is not converted in heat and is related with the reversible bending of the domain walls. It is just an example of how the inadequate use of the Maxwell´s relation can lead to a fake conclusion. Even when the entropy change observed was larger than the observed in Gd based compound it is not possible to use this change in applications. The presence of hysteresis in the magnetization vs magnetic field curves is indicative of the presence of an additional term in the first law of thermodynamics that must be considered before any conclusion. In the studied case, this feature was observed in the FM to PS transition meanwhile it is not present in the FM transition, where the entropy change is converted in a temperature change as expected.

**Acknowledgments.**

This work has been done with the support of ANPCyT PICT 1327/2008, Conicet PIP 00889 and UNSAM SJ10/13. We are grateful with Leticia Granja and Roberto Zysler for fruitful discussion and to Joaquin Sacanell for careful reading of the manuscript. MQ is also member of CIC CONICET.

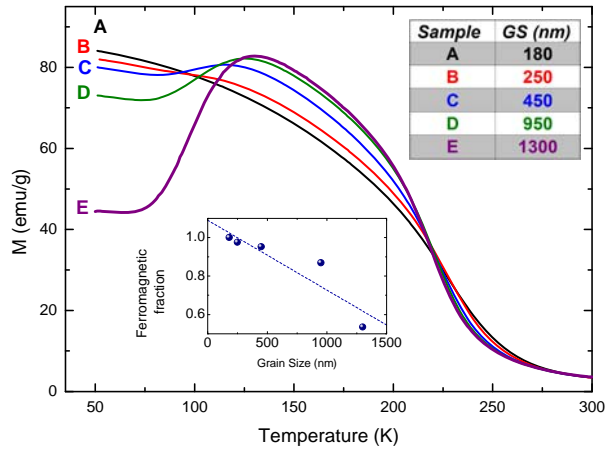

Figure 1: Magnetization as function of temperature with an applied magnetic field of 1 T for samples with different grain size. Inset: ferromagnetic fraction at 50K as function of the grain size. Table: grain size for each sample.

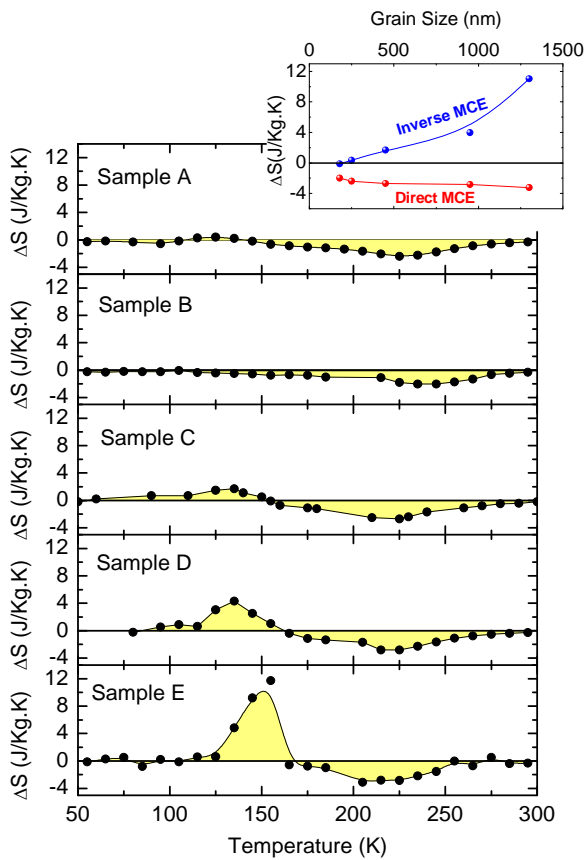

Figure 2: Entropy change as function of temperature for all the samples with a magnetic field of 3 T. b) Intensity of both peaks in the entropy change as a function of GS.

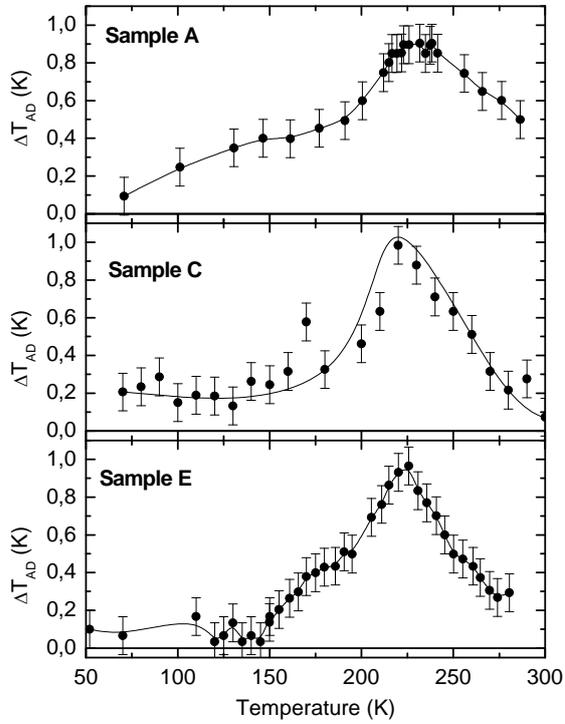

Figure 3: Adiabatic temperature change (ΔT$_{AD}$) for samples A, C and E as function of temperature when the magnetic filed is increased from 0 to 3 T. The values of ΔT$_{AD}$ where extracted from DTA measurements taking into account the heat exchange between the sample and the sample holder.

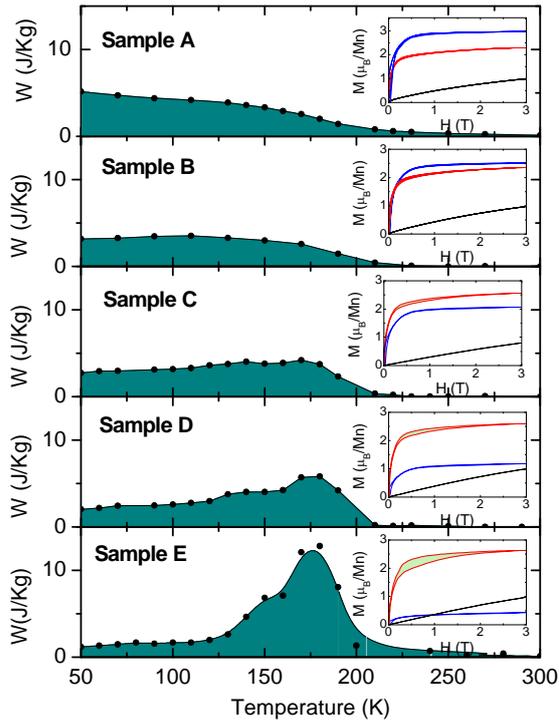

Figure 4: Magnetic work defined as the area enclosed by the increasing and decreacing magnetic field curves as function of temperature for different samples. In the insets we show magnetization loops at 250 K (black), 170 K (red) and 60 K (blue).

[1] V. K. Pecharsky and K. A. Gschneidner, Jr., Phys. Rev. Lett. 78, 4494 (1997).

[2] H. Wada and Y. Tanabe, *Appl. Phys. Lett.* 79, 3302 (2001)

[3] A. Planes, L. Manosa, X. Moya, T. Krenke, M. Acet, and E. F. Wassermann, J. Magn. Magn. Mater. 310, 2767 (2007); C. Salazar Mejia, A. M. Gomes and N. A. de Oliveira, J. Appl. Phys. 111, 07A923 (2012)

[4] Manh-Huong Phan, Seong-Cho Yu, Journal of Magnetism and Magnetic Materials 308 (2007) 325–340; A. Rebello, V. B. Naik, and R. Mahendiran, J. Appl. Phys. 110, 013906 (2011)

[5] T. Krenke, E. Duman, M. Acet, E. F. Wassermann, X. Moya, L. Mañosa, and A. Planes, Nature Mater. 4, 450 (2005); A. M. Gomes, F. Garcia, A. P. Guimaraes, M. S. Reis, and V. S. Amaral, Appl. Phys. Lett. 85, 4974 (2004); P. J. von Ranke, N. A. de Oliveira, B. P. Alho, E. J. R. Plaza, V. S. R. De Sousa, L. Caron, and M. S. Reis, J. Phys.: Condens. Matter 21, 056004 (2009).

[6] M. Balli, D. Fruchart, D. Gignoux, and R. Zach, Appl. Phys. Lett. 95, 072509 (2009).

[7] A. M. Tishin and Y. I. Spichkin, The Magnetocaloric Effect and Its Application (IOP, Bristol and Philadelphia, 2003).

[8] M. Földeàki, R. Chahine and T. K. Bose, J. Appl. Phys. 77, 3528 (1995)

[9] Mohamed Balli, Daniel Fruchart, Damien Gignoux and Ryzard Zach, Appl. Phys. Lett. 95, 072509 (2009); A. de Campos, D. L. Rocco, A. Carvalho, G. Magnus, L. Caron, A. A. Coelho, S. Gama, L. M. D. Silva, F. C. G. Gandra, A. O. D. Santos, L. P. Cardoso, P. J. von Ranke, and N. A. de Oliveira, Nature Mater. 5, 802 (2006).

[10] Weibin Cui, Wei Liu and Zhidong Zhang, Appl. Phys. Lett. 96, 222509 (2010)
11 G. J. Liu, J. R. Sun, J. Shen, B. Gao, H. W. Zhang, F. X. Hu and B. G. Shen, Appl. Phys. Lett. 90, 032507 (2007)

[12] M. Quintero, L. Ghivelder, A. M. Gomes, J. Sacanell and F. Parisi J. Appl. Phys. 112, 103912 (2012)

[13] M. H. Phan and S. C. Yu, J. Magn. Magn. Mater. 308, 325 (2007); Z. B. Guo, Y.W. Du, J. S. Zhu, H. Huang,W. P. Ding, and D. Feng, Phys. Rev. Lett. 78, 1142 (1997);P. Sarkar, P.Mandal, and P.Choudhury, Appl. Phys. Lett. 92, 182506 (2008).

[14] Uehara, M; Mori, S; Chen, CH; Cheong SW, NATURE Vol. 399 6736 P 560-563 (1999).

[15] Tebano, A.; Aruta, C.;Medaglia, P. G.;Tozzi, F.;Balestrino, G.;Sidorenko, A. A.;Allodi, G.;De Renzi, R.;Ghiringhelli, G.;Dallera, C; Braicovich, L.;Brookes, N. B.; Phys. Rev. B, 74, 245116 (2006); Curiale, J.;Sanchez, R. D.;Troiani, H. E.;Ramos, C. A.;Pastoriza, H.;Leyva, A. G.;Levy, P.; Phys. Rev. B 75, 224410 (2007)

[16] P. Levy, F. Parisi, G. Polla, D. Vega, G. Leyva and H. Lanza, Phys. Rev. B 62, 6437 (2000).

[17] E. Dagotto, New J. Phys. 7, 67 (2005).

[18] Colossal magnetoresistance oxides, edited by Y. Tokura, Monographs in Condensed Matter Science (Gordon & Breach, New York, 1999)

[19] P. G. Radaelli, D. E. Cox, M. Marezio and S. W. Cheong, Phys. Rev. B 55, 3015 (1997)

[20] P. G. Radaelli, D. E. Cox, M. Marezio, S. W. Cheong, P. E. Schiffer and A. P. Ramirez, Phys. Rev. Lett. 75, 4488 (1995).

[21] P. Amirzadeh, H. Ahmadvand, P. Kameli, B. Aslibeiki, H. Salamati, A. G. Gamzatov, A. M. Aliev and I. K. Kamilov, J. Appl. Phys. 113, 123904 (2013).

[22] N. S. Bingham, P. Lampen, M. H. Phan, T. D. Hoang, H. D. Chinh, C. L. Zhang, S. W. Cheong, and H. Srikanth, Phys. Rev. B 86, 064420 (2012)

[23] I. G. Deac, S. V. Diaz, B. G. Kim., S. W. Cheong and P. Schiffer, Phys. Rev. B, 65, 174426 (2002).

[24] L. Ghivelder, F. Parisi, Phys. Rev. B 71, 184425 (2005)

[25] J. S. Park,C. O. Kim,Y. P. Lee,Y. S. Lee, H. J. Shin,H. Han and B. W. Lee Journal of Applied Physics 96, 2033 (2004); M. Pekala, V. Drozd, J. F. Fagnard, Ph. Vanderbemden,. J. Alloys and compounds 507, 350-355 (2010)

[26] B. Samantaray, S. K. Srivastava and S. Ravi, Journal of applied Physics 111, 013919 (2012)
[27] J. B. Goodenough, Phys. Rev. 100, 564 (1955)
[28] P. Chen, Y. W. Du and G. Ni, Europhys. Lett., 52 (5), pp. 589-593 (2000)
[29] P. Sande, L. E. Hueso, D. R. Miguéis, J. Rivas, F. Rivadulla and M. A. López-Quintela, Appl. Phys. Lett., 79, 2040 (2001).
[30] Yamila Rotstein Habarnau, Pablo Bergamasco, Joaquin Sacanell, Gabriela Leyva, Cecilia Albornoz, Mariano Quintero, Physica B, volume 407, issue 16, pp. 3305 – 3307 (2012) ; A. Schilling and M. Rebelt, Review of Scientific Intstuments 78, 033904 (2007).

[31] D. Jiles, D. Atherton, J. Magn. Magn. Mater., 61 (1986), pp. 48–60; D. C. Jiles and D. L. Atherton, J. Phys. D: Appl. Phys., 17, 1265-1281 (1984)